\documentclass[conference,usenatbib]{basi}
\usepackage[T1]{fontenc}
\usepackage[british]{babel}
\usepackage[varg]{txfonts}
\usepackage{rotating}
\usepackage{dcolumn}
\begin{document}
\title[Extracting Flow parameters in 2010 outburst of H 1743-322 with TCAF model]
{Extracting Flow parameters of H 1743-322 during early phase of its 2010 outburst using Two Component Advective Flow model}
%
\author[Debnath et al.]%
       {Dipak Debnath$^1$\thanks{email: \texttt{dipak@csp.res.in}},
       Sandip K. Chakrabarti$^{1,2}$, and Santanu Mondal$^{1}$\\
       $^1$Indian Center for Space Physics, 43 Chalantika, Garia St. Rd., Kolkata, 700084, India.\\
       $^2$S. N. Bose National Centre for Basic Sciences, Salt Lake, Kolkata, 700098, India.}

\pubyear{2013}
\volume{**}
\pagerange{**--**}

\date{Received --- ; accepted ---}

\maketitle
\label{firstpage}

\begin{abstract}
We study the spectral properties of Galactic transient black hole candidate 
H~1743-322 during its early phase of 2010 outburst with Two Component Advective Flow (TCAF) model, 
after its inclusion in spectral analysis software package XSPEC as a local model. 
For the analysis, spectral data from RXTE/PCA instrument in $2.5-25$~keV energy band 
are used. From the spectral fit, accretion flow parameters such as Keplerian (disk) rate, 
sub-Keplerian (halo) rate, location of the shock and strength of the shock are directly extracted.
QPO frequencies are predicted from the TCAF model spectral fitted shock parameters, 
`closely' matches with the observed frequencies. 
\end{abstract}

\begin{keywords}
 Black Holes, shock waves, accretion disks, Stars:individual (H~1743-322) 
\end{keywords}

\section{Introduction}

The Galactic transient black hole candidates (BHCs) are interesting objects to study in X-rays
because these sources generally show evolutions in their timing and spectral properties during
their outbursts, which are strongly correlated to each other. In last few decades, more precisely
after the launch of RXTE satellite, our understanding on BHCs have progressed a lot, although still we 
believe that there is a large space to improve our knowledge on the accretion flow dynamics around 
the BHCs. Here, we try to understand this important physical property of H~1743-322 during the early 
phase of its recent 2010 outburst by the spectral study using Two Component Advective Flow (TCAF) 
model \citep[][hereafter CT95]{CT95}.

It is a well known fact that the standard Keplerian disk \citep{SS73} cannot explain everything 
about the X-ray spectrum from BHCs, one necessarily require another flow component, namely 
the `Compton' cloud \citep{ST80}, which produces the non-thermal power-law part of the spectrum. 
In TCAF model, this Compton cloud is replaced by a low angular momentum ({\it sub-Keplerain}) flow. 
Matter in this accretion flow becomes hot close to the black hole, where the centrifugal pressure starts 
dominating and an accretion shock may or may not form depending on whether or not the shock condition is 
satisfied \citep[CT95;][hereafter C97]{C97}.


The recurrent low-mass X-ray binary candidate H~1743-322 has shown several epochs of X-ray flaring 
activity in the last decade. Recently in 2010, it showed X-ray outburst \citep{Yamaoka10} which 
continued for around two and half months and was monitored with RXTE on a daily basis. 
Detailed timing and spectral properties and their evolutions during the 2010 \& 2011 outbursts of 
H~1743-322 are presented in \citet{DD13a}. Evolutions of QPO frequencies as of other transient BHCs 
\citep{C08,C09,DD10,Nandi12} are also observed during the both rising and declining phases of the 
outbursts, fitted with propagating oscillatory shock (POS) solution. In \citet{DD13a}, spectral 
properties are studied with the combination of disk black body and power-law models. 
Here, spectral properties of the source during the early phase of its 2010 outburst are studied 
with TCAF model (CT95), after its inclusion in XSPEC. Also from the spectral fitted shock parameters, 
frequency of the observed QPOs are predicted, which approximately matches. 
The {\it paper}, is organized in the following way. 
In next Section, 
we present observation results based on the TCAF model spectral fit. Finally, 
in \S 3, we present the brief discussion and make some concluding remarks.

%

\section{Observation and Data Analysis}

We present spectral analysis results of publicly available archival data from the RXTE PCA instrument for 
the early phase of the 2010 outburst of H~1743-322. We have extracted and analyzed the RXTE archival spectral 
{\it standard2} mode data starting from $2010$ August $09$ (Modified Julian date, MJD=$55417$) to $2010$ August $16$ 
(MJD=$55424$) from the PCA instrument. For the spectral study using TCAF model, we first generated model {\it fits} 
file by feeding $\sim 4\times 10^5$ numbers of theoretical model spectra in a program written in FORTRAN \citep[for detailed 
properties of the model and its implementation in XSPEC see,][]{DD13b}. The model spectra are generated by varying 
five input parameters in modified CT95 code. 

$2.5-25$ keV PCA background subtracted spectra are fitted with TCAF model {\it fits} file. To achieve best fit
a Gaussian line of peak energy around $6.5$keV is used. For the entire outburst, the hydrogen column density 
(N$_{H}$) is kept fixed at $1.6 \times 10^{22}$ \citep{DD13a} and also $1.5$\% systematic error is assumed. 
For the spectral fittings with TCAF model, one needs to supply five model input parameters: 
$i)$ Keplerian rate ($\dot{m_d}$ in Eddington rate), $ii)$ sub-Keplerian rate ($\dot{m_h}$ in Eddington rate),
$iii)$ black hole mass ($M_{BH}$) in solar mass ($M_\odot$) unit, $iv)$ location of the shock ($r_s$ in Schwarzschild
radius $r_g$=$2GM/c^2$), and $v)$ compression ration ($R$) (ratio between post- and pre- shock densities, i.e., 
equal to $\rho_+ / \rho_-$) of the shock other than model normalization value ($norm$), which is equivalent to 
$\frac{1}{4\pi D^2} cos(i)$, where $D$ is the source distance in $10$~kpc unit and $i$ is the disk inclination angle.

\begin{figure}
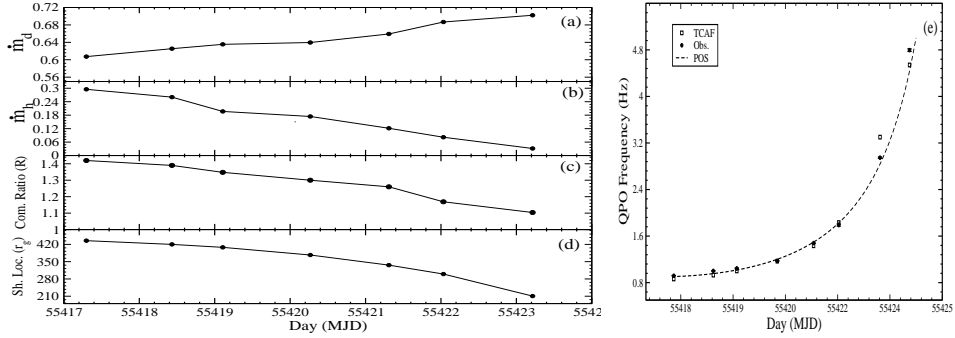

\vbox{
\vskip 0.2cm
\centerline{
\includegraphics[scale=0.6,angle=0,height=4.4truecm,width=8truecm]{fig1.eps}
\includegraphics[scale=0.6,angle=0,height=4.4truecm,width=4.4truecm]{fig2.eps}}
\vspace{0.0cm}
\caption{(a-e) Variation of TCAF model fitted spectral parameters (Keplerian rate $\dot{m_d}$, sub-Keplerian 
rate $\dot{m_h}$, compression ratio $R$, and shock location $r_s$) are shown in the plots of left panel
and in right panel, recalculated QPO frequencies from the TCAF model fitted shock parameters and from the 
observations are shown. Also, day-wise QPOs calculated from POS model solution (dotted curve) as presented 
in \citet{DD13a} are shown in this plot.}}
\end{figure}

In Fig. 1(a-d), the variation of TCAF model fitted spectral parameters, such as two types of accretion flow rates 
($\dot{m_d}$ and $\dot{m_h}$), and shock parameters ($R$ and $r_s$) are discussed. From the figure it is clear 
that as time passed, disk becomes more cooler with the rise in Keplerian component and fall in sub-Keplerian rate.

According to shock oscillation model (SOM), shock wave oscillates either because of a resonance, where the cooling 
time of the flow is approximately the infall time \citep{MSC96} or because the Rankine-Hugoniot condition is not 
satisfied \citep{Ryu97} to form a steady shock and the QPO frequency is inversely proportional to the infall time 
($t_{infall}$) in post-shock region. According to POS (which is nothing but a time varying version of SOM) solution 
\citep{C08,C09,DD10,DD13a,Nandi12}, one can obtain the QPO frequency ($\nu_{QPO}$) if one knows the instantaneous 
shock location compression ratio or vise-versa. Here from the TCAF model spectral fit, these two important shock 
parameters are extracted and from there we have predicted frequency of the observed QPOs as defined in the equation 
$\nu_{QPO}=\frac{C}{R~r_s~(r_s-1)^{1/2}}$, 
where $C$ is a constant, equals to $M_{BH}\times 10^{-5}$, and $M_{BH}$ is the mass of the BHC.
In Fig. 1e, these predicted QPO frequencies along with observed and day-wise calculated POS model 
fitted QPOs \citep{DD13a} are shown. From Fig. 1(c-d), it is clear that with time shock moved inward 
with reducing strength; as a result of which, frequencies of the observed QPOs are increased.

\section{Discussions and Concluding Remarks}

Here, we present preliminary results based on TCAF model spectral fit of the early rising phase 
of 2010 outburst of the Galactic transient BHC H~1743-322. We have included TCAF model 
in HEASARC's spectral analysis software package XSPEC as a local additive model after generation of model 
{\it fits} file using a bank of theoretical spectra. From the spectral study with TCAF model, accretion 
flow properties of the source during the early phase of the outburst are understood in a better manner, 
where variation of two component accretion flow rates as well as shock parameters are observed directly. 

From the spectral fit, it is also clear that initially spectra are dominated with the Comptonized sub-Keplerian 
halo rate and the source is found to be in hard/low-hard state. As time goes, supply from the Keplerian component 
increases, as a result of which, the source moved towards the softer states. 
On the first day of our observation, shock was found to be at $\sim 435~r_g$, and it moved towards the black hole 
horizon with time (day) with decreasing shock strength and reaches $\sim 175~r_g$ on the last day of our 
present observation. From the shock parameters the frequency of the observed QPOs are also predicted, 
which roughly matches with the observed as well as POS model fitted values. 
The evolution of the spectral properties and observed QPO frequencies as predicted from 
the TCAF model spectral fit for 2010 \& 2011 outbursts of H~1743-322 will be presented 
in our future journal articles \citep{SM13,DD13b}.

%


\end{document}